# Quantum Theory and the Observation Problem[*]

Ravi V. Gomatam[†]

ABSTRACT

Quantum theory is applicable, in principle, to both the microscopic and macroscopic realms. It is therefore worthwhile to investigate whether it is possible to evolve a quantum-compatible view of the properties and states of macroscopic objects in everyday thinking. It will allow a realistic interpretation of quantum theory in a manner directly consistent with the observations. The construction of such a view will provide a solution to what I term the "observation problem".

Toward solving the observation problem, I identify a category of new objective properties called 'relational properties' that are (so to speak) in-between primary and secondary properties. We regularly associate such properties with everyday objects, and I discuss how in fact these are quantum-compatible. If this relational-property viewpoint could be worked into quantum theory, it would altogether avoid the measurement problem, which is an artifact of our current inconsistent (albeit pragmatically successful) strategy of retaining a classical view of the macroscopic world while applying quantum theory to the microscopic world.

Some implications of the relational property viewpoint to neurobiological issues underlying cognition are touched upon.

## I: Introduction

> "Every analysis of the conditions of human knowledge must rest on considerations of the character and scope of our means of communication."
>
> **Niels Bohr (1957, p. 88)**

---

How does physics describe reality? Theoretical terms are never directly observable, not even in early mechanics. The speculative character of all scientific concepts, even that of 'position' in early mechanics was adequately emphasized by Mach. Newton himself was only too acutely aware of the formal nature of his theory's 'description':

> I . . . use the word 'attraction', 'impulse', or 'propensity' of any sort toward a centre, promiscuously and indifferently, one for another, considering those forces not physically but mathematically; whereof the reader is not to imagine that by those words I anywhere take upon me to define the kind or the manner of any action, the causes or the physical reason thereof, or that I attribute forces in a true and physical sense to certain centres (which are only mathematical points) when at any time I happen to speak of centres as attracting or as endued with attractive powers (Newton, 1687, Definition VIII, pp. 5-6).

Duhem and Einstein, among others, have additionally pointed out that only the theory as a whole gets verified in experience (Howard, 1990). Furthermore, data under-determines the theory. From these considerations it follows that the pragmatic success of a theory alone cannot give its individual concepts the status of being 'real'. Scientific realism needs further justification. Ordinary language remains the sole source of our notions of the 'real' in everyday thinking. If so, naming theoretical terms using words of the ordinary language would be a necessary starting step for explicating the nexus we conceive between scientific terms and reality. The need for physical theories to connect with ordinary language was eloquently expressed by Pauli:

> It is true that these [*scientific*] laws and our ideas of reality which they presuppose are getting more and more abstract. But for a professional, it is useful to be reminded that behind the technical and mathematical form of the thoughts underlying the laws of nature, there remains always the layer of everyday life with its ordinary language. Science is a systematic refinement of the concepts of everyday life revealing a deeper and, as we shall see, not directly visible reality behind the everyday reality of coloured, noisy things. But it should not be forgotten either that this deeper reality would cease to be an object of physics, different from the objects of pure mathematics and pure speculation, if its links with the realities of everyday life were entirely disconnected (Pauli, 1994, p. 28, italics mine).

Ordinary language notions of the real--apart from helping to visualize the abstract theoretical terms--play another, more fundamental role in physical theory. The predictive content of any physical theory needs to be stated first, not as scientific observations using the formal terms, but as everyday experiences in the laboratory using ordinary language. Such a description of observation experiences using ordinary language is independent of the physical theory, though it must be compatible with the theory. Ordinary language and the realist intuitions underlying it will thus play an important and critical role within scientific realism from the very outset.

Our present view of the everyday world allows for macroscopic objects to factually have only one definite physical state, out of many logical possibilities. If we toss a coin, logically speaking, both heads and tails are possibilities. In an actual toss, however, only heads or tails will show up. We shall refer to this view of the objects of the everyday world as the classically-definite (CDEF) conception of an object and its state. According to this CDEF conception, an object is always (factually) in only one of its many (logically) possible states. In addition, all the properties that we can associate with such a factual state of the object will have determinate values at all times, whether measured or not. This CDEF conception is independent of any physical theory and is basic to everyday naive realism.

In this paper I shall argue the possibility for abandoning the CDEF conceptions altogether while interpreting quantum theory, even at the observational level, despite their demonstrated pragmatic usefulness. I shall argue for this move by raising both theoretical and experimental considerations that call into question how far terms such as wave-particle duality and superposition adequately convey the quantum implications of the corresponding formal terms. Indeed, the central thesis of this paper will be that we cannot even begin to comprehend the essential nature of the quantum mechanical description unless we develop an alternative, quantum-compatible conception of everyday objects in everyday thinking. We shall call the task of identifying such a conception of everyday objects and developing appropriate formal ideas based on it as the 'observation problem'. We then discuss certain interpretive insights of Einstein and Bohr that shows both of them recognized the observation problem as the principal one. Toward solving the observation problem, I identify a range of properties I label as 'relational properties'. We routinely attribute to macroscopic objects in everyday thinking, and I discuss how in fact they are quantum-compatible. I discuss in broad terms how incorporating this relational property viewpoint into quantum physics would solve the observation problem. Such a solution would also simultaneously altogether avoid the measurement problem, which is an artifact of our current pragmatically successful strategy of retaining a classical view of the macroscopic world, while applying quantum theory to the microscopic world.

I conclude by touching upon the possible nexus between the proposed relational property view within quantum physics and the neurobiology of cognition, particularly in the context of intentionality and causality.

## II: The Two-Slit Experiment, the Classically-Definite (CDEF) Conceptions and the Wave-Particle Duality

> "We choose to examine a phenomenon which is impossible, absolutely impossible, to explain in any classical way, and which has in it the heart of quantum mechanics. In reality, it contains the only mystery."
>
> R.P. Feynman (1965, vol. III, p. 1)

Bohr, among all writers on quantum theory, consistently referred to wave-particle duality as an apparent dilemma, a problem to be got rid of, not explained (see, for example, Bohr, 1934/1961, pp. 15, 93, 95, 107). Indeed, our laboratory observations concerning the behavior of quantum objects are always localized, a fact that promotes, if any, a 'particle picture' of the quantum objects. What experimental evidence compels us to ascribe a 'wave nature' to the same objects? A simple step-by-step analysis of the two-slit experiment in several different configurations will show that the so-called wave aspect of the particle at the experimental level is merely an inference resulting from an implicit application of a series of CDEF-based intuitions to the laboratory observations. The basic experimental setup consists of a two- slit screen placed in front of a low-intensity monochromatic light source at an appropriate distance. A battery of detectors (or a photographic plate, effectively an array of detectors) is placed on the other side of the screen.

*Configuration 1*: Both slits of the screen are kept open, and a battery of detectors is placed at any place on the other side of the screen. No matter how small we make the size of the detector, we always see only one detector click.

*CDEF conclusion 1:* Light is always observed to be a localized 'particle'. Let us call them 'photons', following conventional practice.

**Configuration 2**: Both slits of the screen are kept open, and two detectors are placed very close to the screen, one at each slit. Predictably, at any instant only one of the detectors always fires.

*CDEF conclusion 2:* Photons always 'go through' one or the other slit.

**Configuration 3**: The photographic plate is placed far enough from the two-slit-screen (depending on the distance between the slits and the size of the slits), and one slit is closed. The localized light spots continue to appear, but randomly and everywhere on the photographic plate with a peak in one region which is dependent on which slit is open.

*CDEF conclusion 3:* Photons going through either slit can arrive at any spot on the photographic plate.

**Configuration 4**: Same as configuration 3, except that now both slits are kept open. Again, localized spots appear randomly on the photographic plate. However, when we let enough of the spots accumulate, a pattern of alternate dark and bright bands appears on the screen, familiarly called the 'interference' pattern.

*CDEF conclusion 4:* when both slits are open, even though single photons (must) continue to be going through one or the other slit, we now see that there are regions where the photons don't arrive at all!

Now, using CDEF conceptions, one further infers the following: consider a spot P on the photographic plate which falls in the 'dark region' when both slits are open. Of all the single photons that 'go through' the right slit (according to the results of configuration 3), some that would arrive at P if the left slit were closed, do not when the left slit is open!

CDEF conclusion 5: Where a photon going through one of the slits (and all photons must go through one or the other slit, according to the results of configuration 2) would arrive on the screen depends on the status of the other slit (open or closed).

Penrose (1989) has called it the 'most mysterious part of the two-slit experiment'. In some unknown manner, the photon reacts to the status of both slits, and this allows us to ascribe an extended or 'wave' nature to the photon. Dirac has famously suggested that the single photon, going through the slits, somehow interferes with itself. There is also the notion that, prior to an actual laboratory observation the photon has many constituent states that correspond to, in some vague sense, possibilities for going through both slits.

It is important to see that the idea of mysterious 'self-interference' appears necessary only because of the following assumptions:

(a) We accept the 'particle' idea in view of the localized nature of individual observations.

(b) We make further CDEF inferences in analysing the outcomes in each experimental configuration.

(c) We combine these inferences across experiments into a single picture of the atomic world.

## III: The CDEF Conceptoin at the Everyday Level and Its Incompatibility with the Quantum Formalism

At the core of the quantum mechanical description is a wave function, denoted by $\Psi$. Its evolution is governed by the Schrödinger equation. If two wave functions, $\psi_1$ and $\psi_2$ are possible solutions to the Schrödinger equation, then the principle of superposition holds that a linear combination of the two, $\psi = c_1\psi_1 + c_2\psi_2$ is also a solution. However, within a CDEF conception of the world, we find our laboratory observations always correspond to either state $\psi_1$ or $\psi_2$, never both. It would be more congenial if we could devise a non-CDEF conception of the observations that would allow us to relate the laboratory observations to $\Psi$ as well.

However, any such inclusive conception would have to revise, not only our notion of the states of atomic objects, but of macroscopic objects also. This point was best illustrated by Schrödinger in his famous 'cat paradox'. A single photon in a state of superposition is connected to a detector, which is wired up to release poison gas inside a box in which a living cat has been placed. It is possible to visualize, in principle, a wave function that describes the joint state of the atom, counter, poison trigger, all the way up to the cat. Such a wave function would require that, prior to a laboratory observation, the cat also be in a state of superposition, i.e. a state that does not permit any CDEF-based notion. Nevertheless, independent of quantum theory, CDEF notions would tell us that a cat has to be in one of two distinct states, dead or alive, at all times.

We may thus tend to conclude that Schrödinger's paradox cannot point to any problem at the observational level. Instead, we may conclude that the paradox shows the utter necessity to accept that, with quantum theory, the physical descriptions of the world have become purely abstract in the sense that they cannot be brought in contact with our everyday notions of the world, even in an idealized way.[1]

However, the 'cat' in the paradox only serves the role of a macroscopic object, the point of the paradox being QT entails mutually inadmissible possibilities for *physical states* to 'co-exist' in a manner that is manifestly impossible at the macroscopic level. Therefore, the macroscopic object need not be a biological entity. A flag which would either go up or down a pole, depending on whether or not the detector registers the photon can easily replace the cat. Thus, one way out of the 'paradox' is to ask whether the physical states of macroscopic objects can be conceived in terms other than mutually exclusive CDEF states. There is no logical reason to suppose a priori that this is impossible. In fact the aim of this paper is to eventually show one such possible conception, although its actual application within quantum theory must await further work.

Another important consideration is that quantum theory is applicable to macroscopic objects also. Thus, it is not possible to solve Schrödinger's cat paradox by retaining the CDEF notions at the macroscopic level, and limiting the implications of superposition to the atomic level. Yet, all extant interpretations begin by adopting precisely such a move. This immediately introduces the famed 'measurement problem'. We have, on the one hand, a description of atomic objects involving superposition that does not permit any CDEF compatible intuitions. Yet, our observations in relation to these superposed states are always CDEF compatible. At present, various interpretations differ only over how to reconcile these two different conceptions of states, while agreeing on the use of the CDEF notions at the level of observation. In solving the problem, which arises only as a consequence of the commitment to the CDEF view at the level of observation, they seem to inevitably admit a role for the conscious observer to account for the appearance of a classically-definite state. This role is antithetical to the very spirit underlying the CDEF conceptions they assume to start with.

For example, in the 'collapse models', the superposed state is taken to be an objectively real state of the quantum system in between measurements, while a physical non-local mechanism is taken to resolve superposed state into a CDEF compatible observed state. Indeed, some physicists have invoked the objective reduction of the collapse-model to link quantum theory with consciousness. Some recent examples are Penrose (1989); Goswami (1989; 1993); Stapp (1993). Another example of a realist interpretation is the 'many-worlds interpretation', in which the state of superposition represents a collection of worlds in each of which the CDEF conception holds. Why does one particular result that we observe take place in our world? The interpretation can only account for it as a fact of our experience. Thus, this interpretation too requires a role for our subjective consciousness.

## IV: On the Need for an Alternate Non-CDEF Conception of the Everyday World

What grounds do we have for assuming that within quantum theory CDEF notions apply at the level of observations, when quantum theory applies in principle to both microscopic and macroscopic objects? The correspondence principle is taken to provide the justification here. According to this principle, in the limit in which the Planck's constant is negligible compared to the quantities being measured, and this is generally the level of macroscopic bodies, the quantum mechanical description reduces, for all practical purposes, to that of classical description. However, Bohr himself wrote the following with reference to the correspondence principle:

> In the limit of large quantum numbers . . . mechanical pictures of electronic motion may be rationally utilized. It must be emphasized, however, that this connection cannot be regarded as a gradual transition towards classical theory in the sense that the quantum postulate would lose its significance for high quantum numbers. On the contrary, the conclusions obtained from the correspondence principle with the aid of classical pictures depend just upon the assumptions that the conception of stationary states and of individual transition processes are maintained even in this limit (1934/1961, p. 85).

Einstein too insisted that the fact that quantum theory requires macroscopic objects to be in principle in the state of superposition is significant for the issue of interpretation (1969/1949, p. 682).

The practical successes in applying quantum theory using the CDEF conceptions of the observations in the everyday world (via the correspondence principle) has promoted the rather pervasive conclusion that the radical implications of quantum theory are limited to the atomic realm. Our point of departure is to propose that the 'cat paradox' points to the need to go back and devise a non-CDEF, quantum- compatible conception of everyday objects for interpreting the observations. Schrödinger himself, in his later years, wrote along lines compatible with such an understanding of the cat paradox:

> It is probably justified in requiring a transformation of the image of the real world as it has been constructed in the last 300 years, since the re-awakening of physics, based on the discovery of Galileo and Newton that bodies determine each other's accelerations. That was taken into account in that we interpreted the velocity as well as the position as instantaneous properties of anything real. That worked for a while. And now it seems to work no longer. One must therefore go back 300 years and reflect on how one could have proceeded differently at that time, and how the whole subsequent development would then be modified. No wonder that puts us into boundless confusion! (Schrödinger, Letter to Einstein, 18 November 1950, in Prizibram, 1967, p. 38)

Schrödinger is raising the need to go back and redo the physics of macroscopic objects. This is compatible with envisioning the need for an alternative to the CDEF conception of observations and associated macroscopic objects to realistically interpret quantum theory.

Let us allow for the possibility that quantum theory requires an alternative, non-CDEF conception of macroscopic objects. Such a conception would have to be complementary to the present CDEF conception of macroscopic objects. This means that while only quantum conceptions would apply at the atomic level (since classical theory is known to fail at this level), we can choose between two conceptions, one classical and another quantum, at the observational level. This could explain why, while having unmitigated success in using quantum theory practically by choosing to treat the everyday world within the CDEF conceptions, we are nevertheless unable to get an understanding of the state of superposition.

If we are right, in order to consistently interpret quantum theory, the first step would be to discover, in ordinary language, words to describe the states of macroscopic objects that are compatible with the formal principle of superposition. We shall designate the task of finding such a description of macroscopic objects as the 'observation problem'.

The observation problem aims to realistically interpret laboratory observations in a quantum-compatible manner using the everyday language. The measurement problem aims to realistically interpret the states of atomic objects directly, using quantum theoretical language. Within the CDEF framework, the task of solving the measurement problem remains highly problematic. To develop an alternative to the CDEF conception requires recognizing the observation problem.

The 'measurement problem' presupposes the 'quantum/classical dichotomy' with measurement as the sole link between the two worlds. The 'observation problem' begins simply by asking what it is that we are observing in the laboratory, if we treat them as simple everyday experiences. We thus see the 'observation problem' as being logically prior to the 'measurement problem'. The measurement problem might well dissolve once the observation problem is recognized and appropriately solved.

To take the observation problem seriously is to expect to go beyond describing the observations simply as meter pointer positions, detector clicks or localized spots on photographic plates. Einstein made a similar point:

> We are like a juvenile learner at the piano, just relating one note to that which immediately precedes or follows. To an extent this may be very well when one is dealing with very simple and primitive compositions; but it will not do for the interpretation of a

> Bach Fugue. Quantum physics has presented us with very complex processes and to meet them we must further enlarge and refine our concept of causality (Einstein, 1931, p. 203).

Einstein is certainly talking about enhancing our notion of causality within physics. However, he developed at length a philosophy of science in which he said the following:

> The whole of science is nothing more than a refinement of everyday thinking. It is for this reason that the critical thinking of the physicist cannot possibly be restricted to the examination of the concepts of his own specific field. He cannot proceed without considering critically a much more difficult problem, the problem of analysing the nature of everyday thinking (1936, p. 349).

Einstein evidently expected the refinement of our notion of causality to proceed from the level of everyday thinking. Indeed, Einstein constantly emphasized the 'freely created nature' of all of our concepts.

Even Bohr, who strongly insisted on the permanency of the CDEF conceptions recognized that any new concepts must first be shown to be applicable in the world of our experience. Only thus could he write so strongly:

> It would be a misconception to believe that the difficulties of the atomic theory may be evaded by eventually replacing the concepts of classical physics by new conceptual forms . . . the recognition of the limitation of our forms of perception by no means implies that we can dispense with our customary ideas or their direct verbal expressions when reducing our sense impressions to order (Bohr, 1934/1961, p. 16).

The foregoing discussion brings out the striking difference between the approaches of Bohr and Einstein. It would seem that, contrary to popular opinion, Bohr was far more classical than Einstein was. Jammer has remarked,

> contrary to widespread opinion, [Einstein] rejected the theory not because he, Einstein - owing perhaps to intellectual inertia or senility - was too conservative to adapt himself to new and unconventional modes of thought, but on the contrary, because the theory was in his view too conservative to cope with the newly discovered empirical data (1982, p. 60).

Nevertheless, both Bohr and Einstein, in contra-distinction to the rest of the physics community, devised their interpretations starting from the common point that quantum theory is incompatible with CDEF-notions even at the level of everyday experience.

## V: Bohr, Einstein and the Observation Problem

If we logically allow for an alternative, quantum-compatible conception of everyday objects, then until we devise such a conception, the observations pertaining to quantum theory should be treated as the experiences we have in the laboratory. Both Bohr and Einstein started from this point[2], and thus could be regarded as having taken the first step in recognizing and solving the observation problem.

> For our theme, however, the decisive point is that the physical content of quantum mechanics is exhausted by its power to formulate statistical laws governing observations

> obtained under conditions specified in plain language (Bohr, 1963, p. 12, emphasis mine; see also p. 61).

> The de Broglie-Schrödinger wave fields were not to be interpreted as a mathematical description of how an event actually takes place in time and space, though, of course, they have reference to such an event. Rather they are a mathematical description of what we can actually know about the system. They serve only to make statistical statements and predictions of results of all measurements, which we can carry out upon the system (Einstein, 1940, p. 491).

Both, however, acknowledged that individual quantum objects are real and cause individual events, based on direct experimental evidence (Bohr, 1934/1961, pp. 100, 102, 112; 1957, pp. 16, 24, 73, 87).

How did they reconcile their realist commitments with their apparently instrumental view of the formalism? Both concluded that more than just the individual system was underlying the laboratory observations of quantum theory.

For Bohr, the system under observation and the experimental means for observing them formed, as Teller put it, a single epistemic whole:

From the beginning, the attitude towards the apparent paradoxes in quantum theory was characterized by the emphasis on the features of wholeness in the elementary processes, connected with the quantum of action. The element of wholeness [has] the consequence that, in the study of quantum processes, any experimental inquiry implies an interaction between the atomic object and the measuring tools which, although essential for the characterization of the phenomena, evades a separate account . . . (Bohr, 1963, pp. 78, 60).

For Einstein, the quantum mechanical Ψ-function captured the state of the ensemble as a 'single whole':

> . . . the Ψ-function does not, in any sense, describe the state of one single system. The Schrödinger equation determines the time variations which are experienced by the ensemble of systems which may exist with or without external action on the single system . . . [quantum mechanics] does not operate with the single system, but with a totality of systems . . . (Einstein, 1936, pp. 375-6).

This too must be necessarily an epistemic idea, because the ensemble of quantum systems can be prepared and observed one at a time in an experiment.

Thus, according to both, the 'cause' underlying each laboratory observation involves more than just the behaviour of the individual system, whose reality is not doubted. Thus, neither Bohr nor Einstein can be called Machian sensationists or instrumentalists of any kind, because of their realist commitment. However, in the absence of an alternative conception, they were obliged to treat their respective holisms epistemologically in order to avoid explicit contradictions with the CDEF-framework. Since their interpretations represent, in our view, the first step toward recognizing the observation problem, we shall discuss some relevant details of their respective interpretations. Both of their interpretations are quite complex, and the present author has treated them in detail elsewhere (Gomatam, 1999).[3] For the present paper, their interpretive ideas will be traced only at a summary level.

# VI: Einstein and the Ensemble Holism

> "Physicists would understand me a hundred years after my time."
>
> **Einstein quoted in Pais (1992, p. 467)**

Einstein's interpretation is an attempt to justify the assumption of the existence of the atomic object as a localized individual whose description is missing within quantum theory. The interpretation, at the same time, aims to account for the theory's pragmatic successes in describing the observable consequences of the behavior of an ensemble of these localized, individual systems.

It is the situation in any probabilistic description that each observation can go only toward verifying the behavior of the ensemble of identically prepared systems. Classical statistical descriptions are compatible with presuming a causal link between each individual observation and the real state of each individual system. However, quantum theory is irreducibly statistical. In this situation, Einstein took the laboratory observations as a whole to be related to the state of the ensemble as a whole. However, 'the programmatic aim of all physics [is] the complete description of any (individual) real situation (as it supposedly exists irrespective of any act of observation or substantiation)' (Einstein, 1969, p. 667). The view that quantum theory is incomplete follows: One would very much like to say the following:

> Ψ stands in a one-to-one correspondence with the real state of the real system . . . if this works, I talk about a complete description of reality by the theory. However, if such an interpretation doesn't work out, then I call the theoretical description 'incomplete' (Letter to Schrödinger, June 19, 1935; cited in Fine, 1986, p. 71).

Einstein held that 'quantum physics deals with only aggregations, and its laws are for crowds, and not for individuals' (Einstein and Infeld, 1966/1938, p. 286). Yet, he also emphasized that there are quantum individuals, and that they cause the individual laboratory observations (1936, p. 337). It would seem that the holist conception of the ensemble that Einstein is arguing for can be thought of more as a mob than a collection of distinct individuals. In a mob, the individuals act without individual identity, as a part of the mob. Such a view of the individual would relate well with the fact that the CDEF notion of the individuals as localized and separable entities fails within quantum theory. However, the failure of locality and separability may be only symptomatic. More fundamentally, the very idea of conceiving the quantum individuals within the space-time continuum may be at fault: 'the problem seems to me how one can formulate statements about a discontinuum without calling upon a continuum (space-time) as an aid' (Letter to Walter Dllenbach, November 1916, Item 9-072 cited in Stachel, 1986, p. 379). Einstein made the same remark again, in 1954, to David Bohm (ibid., p. 380).[4]

Thus, the need for a non-CDEF conception of the individual object within quantum theory suggests itself in the context of Einstein's ensemble holism.[5] Indeed, given that quantum theory applies to macroscopic objects also, Einstein's ensemble holism is not incompatible with the demand for a new conception of everyday objects. Einstein however acknowledged his failure in general to develop new concepts for quantum theory (ibid., p. 380). He was thus forced to take an essentially negative stand, about the impossibility of treating quantum theory as a complete theory of the individual qua individual in the CDEF sense.

Bohr, on the other hand, worked within the prevailing CDEF conceptions of macroscopic objects, systematically interpreting all the non-CDEF features of the quantum formalism as indicating the epistemological limits within which these CDEF conceptions could be invoked to describe laboratory observations. In thus being less ambitious, Bohr's interpretation could form the natural next step in the search for a positive interpretation of quantum theory.

## VII: Bohr and Quantum Relationality

Bohr envisions that within quantum theory 'an independent reality in the ordinary physical sense can neither be ascribed to the phenomena nor to the agencies of observation' (Bohr, 1934/1961, p. 54). This 'inseparability hypothesis' is central to Bohr's interpretation and distinguishes it from the standard textbook interpretation, often referred to as the Copenhagen Interpretation. (Gomatam, 2007) The inseparability hypothesis allows Bohr to treat the laboratory observations as simple experiences and avoid contradiction with CDEF notions within quantum theory. It simultaneously renders the formalism a symbolic procedure, such that the theoretical concepts cannot be directly given physical significance (Bohr, 1963, p. 61). As a result, it is 'the application of the [classical] concepts alone that makes it possible to relate the symbolism of the quantum theory to the data of experience' (Bohr, 1934/1961, p. 16).

Consider a two-slit screen placed in the path of a thermionic source emitting a beam of mono-energetic electrons. A simple analysis of the experiment using classical intuitions shows that we can conceive of two measurable properties of the particle in relation to the experimental arrangement: through which slit the particle would pass, and in which direction it would emerge from the screen. The former can be called 'position at the slits' and the latter, 'momentum at the slits'. Classical intuition also suggests that by placing two detectors close to the slits, we can gain information as to the position of the particle at the screen (also called 'which-path information'). By placing a battery of detectors far away, we could ascertain the momentum of the particle at the screen. We cannot, however, do both with optimum precision at the same time. Therefore, the better our position set-up is, the worse will be our momentum set-up; and vice versa. This straightaway suggests the non-visualizable nature of the quantum formalism, and Bohr repeatedly emphasized the failure of pictures of trajectories of the particles in space and time within quantum theory. It is equally important to realize that the properties of the particle are defined here not absolutely, but in relation to the experimental arrangement we have made.

Bohr's key interpretive response to this situation is to include the experimental arrangement as a part of the very definition of the property under measurement. Thus, the quantum mechanical states express *'relations* between the system and some appropriate measuring device' (Feyerabend, 1961, p. 372, italics added; see also Jammer, 1974, p. 197). Using quantum theory, we interpret each observation as measuring a property that is defined in terms of the spatio-temporal relation between the observed system and the experimental arrangement. Two essentially different experimental configurations (the detectors being placed near or far) let the observed system enter into a 'position' or 'momentum' relation with the experimental arrangement. Bohr concludes that we should never expect to simultaneously define both properties to arbitrary precision in one experiment. Nor should we expect to be able to combine the results obtained from different individual experiments (Bohr, 1963, p. 4).

Bohr's emphasis, then, is on definition, not measurement. Whereas he usual view of the uncertainty relations is that canonical conjugates such as position and momentum cannot be simultaneously *measured* to arbitrary accuracy, Bohr, however, emphasizes the limits to the *definability* of the two properties in question, thus treating the non-commutation relations entirely epistemologically, as undefinability relations, rather than as uncertainty relations:

> Quantum mechanics speaks neither of particles, the positions and velocities of which exist but cannot be accurately observed, nor of particles with indefinite positions and velocities. Rather, it speaks of experimental arrangements in the description of which the expressions 'position of a particle' and 'velocity of a particle' can never be employed simultaneously (Niels Bohr, Second International Congress for the Unity of Science, Copenhagen, June 21-26, 1936).

No doubt, even in pre-quantum theories, the properties are relational. The length of an object, for example, is expressed as a relation between the object and a standard scale. Or, its position is expressed in relation to a chosen co-ordinate reference frame. However, there is a clear difference between the relationality of pre-quantum physics, and the relationality of quantum theory that Bohr's interpretation as we read him implies. In classical theories, a property was expressed by a relation, but inhered in the object. In quantum theory, the property is both expressed by and inhering in the relation. The claim of Bohr seems to be that without actually setting the detectors in place physically, we cannot even define what property is being measured for the quantum object, and as soon as we consummate the relation by an actual laboratory observation, the property is asserted to exist. Thus, the experimental arrangement seems to both create the context necessary for speaking of a property of the object in real terms, and to measure the property.

Bohr's move allows for the introduction of a new quantum notion of property involving relationality to interpret quantum theory. But Bohr himself did not go that route. He instead treated his insight epistemologically, by arguing that the 'inseparability hypothesis' which leads to the relational mode of description, is a result of the limitations of ordinary language we have run into, in the context of quantum theory (Bohr, 1957, p. 25).

Why didn't Bohr attempt an interpretation of quantum theory directly in terms of new 'relational properties'? Bohr steadfastly denied, for some reason, the possibility for devising new non-classical descriptions of everyday objects. He held that 'all new experience makes its appearance within the frame of our customary points of view and forms of perception' (1934/1961, p. 1), and limited such points of view to classical conceptions, i.e. those taken over in pre-quantum theories (see, for example, 1934/1961, p. 16).

Bohr recognized, I believe, that in order to treat it physically, the idea of quantum relationality would have to be first applied at the level of everyday objects:

> Once at afternoon tea in the Institute, Teller tried to explain to Bohr why he thought Bohr was wrong in thinking that the historical set-up of classical concepts would forever dominate our way of expressing our sense experience. Bohr listened with closed eyes and finally only said: 'Oh, I understand. You might as well say that we are not sitting here, drinking tea, but that we are just dreaming all that.' (Bastin, 1971, p. 27).

In the next section we shall present a conception of properties of everyday objects that we routinely invoke in everyday thinking that could provide a point of departure from Bohr's epistemological perspective, to treat his quantum relationality physically.

# VIII: The Notion of Relational Properties

> The whole of science is nothing more than a refinement of every day thinking. It is for this reason that the critical thinking of the physicist cannot possibly be restricted to the examination of the concepts of his own specific field. He cannot proceed without considering critically a much more difficult problem, the problem of analyzing the nature of everyday thinking.
>
> ***Einstein, 1936, p. 349***
>
> I believe that the first step in the setting of a 'real external world' is the formation of the concept of bodily objects and of bodily objects of various kinds.
>
> ***Ibid., p. 350***

We can identify at least three types of properties that we associate with an everyday object: primary, secondary and subjective properties. 'Length' and 'position' are examples of primary properties. An object being 'blue' or 'my father's gift' are examples of secondary and subjective properties, respectively. Ultimately all properties are subjective in the sense they are our notions of objects. Nevertheless, some properties can be treated as subject-independent more than others. A primary property, for example, can be thought of as a property that will inhere in the object even if all conscious observers were to cease to exist in the universe. By contrast, an object will not have a secondary or subjective property unless at least one conscious observer exists within the universe and perceives the object in that manner.

We can operationalize the above distinction between the three properties in a simple way. 'Length' is an observer-independent property because it can be expressed solely in relation to the length of another physical object, the length scale. The existence of a secondary property in an object would ultimately require appealing to the fact of experience of *any one* conscious observer. For this reason, color, although treated to some extent as an objective property in physics, continues to be considered a 'secondary property'. The existence of a subjective property would require appealing to *a particular* conscious observer.

Following Locke, we have generally presumed that physics should deal only with primary properties to remain objective. Indeed, the formal properties of classical physics are idealizations of primary properties. Progress in physics is often times gauged by how far aspects of non-primary properties can be treated objectively, i.e. as primary properties. However, we discussed in section III the problems in treating the properties associated with the quantum mechanical observables as objective properties in the sense of existing in nature with pre-determined values independent of measurement. Indeed, if primary properties were the only conception of objective properties possible, a subjectivity unavoidably enters quantum theory. It is therefore sometimes wondered whether quantum theory signals the end of the Cartesian divide between mind and matter.

However, the statistical predictions of quantum theory in any given experimental set up cannot be in any way changed by a conscious observer. This indicates strongly the ultimately observer-independent status of quantum mechanical properties. We can then certainly ask: are primary properties the only type of objective properties we can conceive of? We saw that Bohr strongly emphasized the relational mode of quantum mechanical description. Could it be that quantum mechanical properties are relational, yet 'objective' properties? In the reminder of this section, we shall identify a category of properties that we routinely associate with everyday objects, which are indeed objective in that they too are expressed in relation to another object, but differ from primary properties in important ways.

Let us consider a macroscopic object, say a book. From the viewpoint of classical mechanics, such an object would be described by a set of numbers representing mass, position, velocity, etc. Since classical theory is deterministic, these numbers are presumed to exhaust all causally relevant properties of the object. Characterizations of the object as a book, or as a 'gift from my father' are taken to be epiphenomenal descriptions, extraneous to physics.

However, in a consistent interpretation of quantum theory, if the macroscopic world would also have a quantum description, then a macroscopic object must possess more physical properties than those accounted for by classical mechanics. Again, an empathetic consideration of ordinary experience suggests some starting clues. A book can be used as a paperweight or a doorstop. Normally, these would be regarded as different uses of an object. However, they can be considered as involving a new kind of objective property, namely a relational property.

The use of the book as a paperweight requires setting it up in a particular spatio-temporal relation with another physical object, i.e. placing the book on top of a stack of papers. Thus 'paperweight-ness' can be regarded as a potential property of the object, which becomes 'physically real' only when the object is placed in an appropriate spatio-temporal relation with another object. Yet, the property itself objectively belongs to the book. Let us call such a property a 'relational property'. It is similar to and yet different from primary properties that physics has so far studied. Both primary and relational properties are objective in the sense that both are defined in relation to another object, and thus can be said to exist in the object independent of the existence of conscious observers. However, as the ensuing discussion will try to show, the two are different in the sense that while the primary properties are only *expressed by* a relation with another object (such as a scale or a clock), the relational properties are expressed by and *actualized* in a relation with another object.

The classical deterministic description indicates a causal closure in terms of primary properties. For relational properties to be not epiphenomenal, we can envisage the proposed relational properties to be complementary to the primary properties of classical mechanics. Since an 'object' in physics, as Eddington put it, is a conceptual carrier of all of its properties, the two sets of properties would provide two different and complementary physical conceptions of the macroscopic object. The 'causal powers' of the macroscopic object conceived in terms of the relational properties should also be then different from and complementary to the causal powers of the same object conceived in terms of its primary properties.

The term 'complementarity' is used here in the sense of two non-intersecting conceptions, which taken together, provide a more complete conception of the object. This also marks the starting point of Bohr's conception of complementarity. However, within quantum mechanics, Bohr introduced the idea of 'complementarity' epistemologically, as a framework for invoking classical properties. We seek to introduce complementarity ontologically, with respect to relational and primary properties. Bohr's complementarity operated with respect to atomic objects, whereas ours pertains to macroscopic objects. Bohr presented the complementary framework as 'a rational generalization of the causal space-time description of classical physics' (1934, p. 87). We are envisioning that the proposed relational property view of the macroscopic object to engender a non-classical notion of causality.

We now reconsider Schrödinger's cat paradox (section III), from the viewpoint of ontological relational properties within quantum mechanics. A single electron in a state of superposition of one of two possible states (spin up/down; $\Psi = \sum_{i=1,2} c_i \psi_i$) is wired up to a flag on a pole, such that a detection mechanism will send the flag either up or down the pole. The central issue is whether the general superposed state $\Psi$ can be conceived so that the fact that the macroscopic measuring device (flag) can take only one of two mutually exclusive CDEF states (up or down) does not pose a problem for interpretation.

We saw that an object can have many relational properties potentially. The key implication, for our present purposes, is that when we actualize one of these properties (by bringing about the necessary spatio-temporal relationship), others will remain potential. Placing the book in a particular spatio-temporal relation with a stack of papers actualizes its 'paperweight-ness', *while still leaving the other property potential*. In the case of the quantum thought-experiment under consideration, if we let 'spin-up' and 'spin-down' be relational properties, then our interpretation of the observations would have to change. Instead of describing the observation as the flag being either up <u>or</u> down the pole (which would make the corresponding properties absolute properties), we would say that the observation reveals which of the two possible *spatio-temporal relations* the system under observation has entered with the measurement device. In such a relational description, the actualization of one of the properties still leaves the other property as potential *in the observed system*. If the observation were to correspond to the spin-up eigenstate, the claim is that the spin-down eigenstate still exists, in some sense, as a potential relational state.

In current theory, the probability amplitude for the non-occurring eigenstate is taken to be zero since $\int_{v}^{v+dv} \psi \psi^* dv$ is taken to give directly the probability of finding a particle in the unit volume $dv$. We instead propose to treat the integral as providing the probability that a detector placed in the localized region $dv$ will register a detection event.[6] Since this disconnects the appearance of the eigenstate from the probability statements, the probabilities for the non-occurring events would reduce to zero at the point of an actual measurement, but the corresponding probability amplitudes would not become zero. According to the Schrödinger equation, the probability amplitude has a nonzero value. From the relational viewpoint, the non-occurring eigenstates remain potential and could be causally efficacious.[7]

The full incorporation of the above relational view within quantum physics will require further work, starting with a philosophy of ordinary language appropriate for quantum theory (see Gomatam, 2004). The relational property view holds, we believe, the resources for allowing a different use of the formalism complementary to the present one, one that takes into account the objectively real existence of the probability amplitudes of the non-occurring eigenstates at the point of observation. The present paper is limited to arguing a redefinition of the problem of interpretation of quantum theory by considering underlying philosophical issues. We have argued for replacing the 'measurement problem' by the 'observation problem', and have discussed, in broad terms, the implications of the relational approach as a possible framework for solving the latter.  It is beyond the scope of the present paper to expand on all the possibilities of the relational approach. Many more of our classical prejudices concerning objects, their states and properties may likely have to be jettisoned before the relational viewpoint can be fully incorporated within physics.  As part of an ongoing effort in this direction, I have discussed elsewhere (Gomatam, 1999) the possibility for introducing 'information' as a basic physical notion within quantum theory.

## IX: The Relational Viewpoint within Physics and the Neurobiological issues Underlying Cognition

According to the dominant trend in current cognitive science known as 'cognitivism', cognition is representational and best explained using the computational framework. Cognitive processes are treated as 'passive', and determined by sensory stimuli originating in the external world.  'Questions of how the brain can a priori create its own goals and then find the appropriate search images in its memory banks are not well handled by cognitivism', since it ignores 'the possible "constructive" role that our brains might play in interpreting the very material of the external stimuli' (Freeman, 1998).  Freeman proposes alternatively that perception is an active process, involving the occurrence of events internal to the brain that precede and shape our very perception of the external stimuli.  Besides assigning 'an internal origin to the constructs that constitute meaning in the brain and that are the basis of effective action into the world', he additionally proposes, correctly in our opinion, that 'the values for these internally generated constructs are in the success or failure of the <u>*relation*</u> of the organism to its environment, not in its brain' (*Ibid*., emphasis mine).  Freeman notes that at present the only idea that comes closest to this view of cognition is the philosophical idea of intentionality in the sense of Brentano (Freeman, 1996, p. 176; Freeman, 1999, section 2).

The 'relational view' we have outlined above to interpret quantum theory contains, interestingly, a *physical* idea with the same characteristic.  A relational property is defined as involving a spatio-temporal relation with another macroscopic object. The property exists potentially in the object regardless of whether the other object actually exists or not.  In this sense, a macroscopic object conceived relationally contains an 'about-ness' to other objects of the world which, when treated as a physical property, can be causally efficacious.

Neuroscience is based on an understanding of the underlying physical and chemical processes in the brain. Presently our understanding of these processes is largely classical. Quantum theory under the relational viewpoint could make it possible to view the brain *as a quantum (i.e. relational) macroscopic object*, having new causal powers in virtue of new physical properties it has, now in *relation* to the environment. Such relational properties, by their very nature, would not be a fixed list of properties of the object.  They would come about and vanish with the changing (spatio-temporal) relations of the object with its environment. This feature of relational properties is very much in line with the felt need in neuroscience to identify physical processes originating in the brain which carry their meaning and causal efficacy in relation to the environment.

## X: Conclusion

Relativity theory, despite its far reaching idea of the space-time continuum, did not oblige us to change our ways of thinking about space and time at the level of practical living in the everyday world. If quantum theory, as argued in this paper, obliges us to actively reconstruct, *in everyday thinking*, our notions of objects and their causal properties in terms of their mutual spatio-temporal relations, it would represent a much greater and profounder impact on human thinking than all previous physical theories.

## Acknowledgements

The author wishes to thank Alan Sommerer and Greg Anderson for many hours of valuable, critical discussions; Alan Sommerer, Greg Anderson, Omduth Coceal, Saul-Paul Sirag and Jean Burns for carefully reading and commenting upon successive versions of this paper; Henry Stapp and Edward MacKinnon for encouraging comments on a preliminary draft; an anonymous reviewer and John Smythies for comments on the concluding sections; Professor Walter Freeman for valuable encouragement and support.

---

[1] "All of direct, human experience and of human intuition applies to large objects. . . . but things on a small scale just do not act that way. So we have to learn about them in a sort of abstract or imaginative fashion, not by connection with our direct experience." (The Feynman Lectures, Vol. III, p. 1)

[2] Stapp (1993) must be credited with emphasizing this aspect in Bohr's interpretation.

[3] The modern-day Copenhagen interpretation is generally taken to have evolved from Bohr's interpretive ideas. However, in most versions, the Copenhagen interpretation interprets the quantum formalism directly and objectively, as a description of the physics of individual quantum systems in time and space. We have already suggested above that both Bohr and Einstein took an instrumental view of the formalism. I have argued elsewhere (Gomatam, 1998) that although the two differed between themselves, if the version of the Copenhagen interpretation that treats the Y-function objectively real is placed on one side of the so-called 'Bohr-Einstein debate', then both Bohr and Einstein would be on the other side.

[4] Stachel's paper is an excellent source and survey of many of Einstein's unpublished ideas in regard to quantum theory.

[5] It might be worthwhile to briefly point out that we see the 'ensemble holism' underlying Einstein's interpretation as being different from and logically prior to the holism due to 'quantum entanglement' that is often discussed in the literature. For a recent discussion of entanglement holism, see Esfeld, 1999. The entanglement-holism presumes both the CDEF-interpretation of relevant laboratory observations and the individuality of the entangled particles, neither of which, as we have amply argued, is justified within quantum theory. The ensemble holism of Einstein instead focuses on the perceived missing identity of a single individual within quantum theory. If so, the physical meaning of entanglement of individual particles may yet depend upon ascertaining first the nature of the individual as described by quantum theory.

[6] Bohr made much the same interpretive move, without stating it as such: 'The physical content of quantum mechanics is exhausted by its power to formulate statistical laws governing observations obtained under conditions specified in plain language' (1963, p.12). Bohr saw this instrumental view of the formalism as a permanent necessity, assuming that 'as a matter of course, all new experience makes its appearance within the frame of our customary points of view and forms of perception' (1961, p. 1). We however introduce this move *provisionally*, arguing the need and the possibility for constructing new, quantum-compatible forms of perception to interpret everyday experience.

[7] I thank B. Josephson for reminding me that the idea of non-occurring potentialities remaining non-zero at the point of an observation is also to be found in the de Broglie-Bohm-Vigier type of hidden-variable theories. However, the 'empty waves' of the hidden-variable theories are denizens of the 3N-dimensional configuration space, while the conception of relational properties (involving a spatio-temporal relation) allows for the objective amplitudes (i.e. potentialities) to have a physical basis in the three dimensional space.